\documentclass[preprint,nofootinbib,aps,superscriptaddress,eqsecnum]{revtex4}
\usepackage{axodraw}

\newcommand{\beq}{\begin{equation}}
\newcommand{\eeq}{\end{equation}}
\newcommand{\beqa}{\begin{eqnarray}}
\newcommand{\eeqa}{\end{eqnarray}}
\def\bea{\begin{eqnarray}}
\def\eea{\end{eqnarray}}

\def\lam{\lambda}

\def\ifmath#1{\relax\ifmmode #1\else $#1$\fi}
\def\half{\ifmath{{\textstyle{1 \over 2}}}}

\def\eq#1{Eq.~(\ref{#1})}

\def\eqs#1#2{Eqs.~(\ref{#1}) and (\ref{#2})}

\def\sqhalf{\ifmath{{\textstyle{1 \over \sqrt{2}}}}}
\newcommand{\snu}{\tilde \nu}

\def\cw{c_W}
\def\sw{s_W}

\def\lp{\lambda'}

\begin{document}

\preprint{\vbox{  
\hbox{}
\hbox{}
\hbox{}
\hbox{}
\hbox{SLAC-PUB-10253} 
\hbox{hep-ph/0311310}
\hbox{November 2003}
}}

\vspace*{3cm}

\title{Neutrino masses in R-parity violating supersymmetric models}

\author{Yuval Grossman}\email{yuvalg@physics.technion.ac.il}
\affiliation{Stanford Linear Accelerator Center, \\
  Stanford University, Stanford, CA 94309 \vspace*{6pt}}
\affiliation{Santa Cruz Institute for Particle Physics, \\
  University of California, Santa Cruz, CA 95064  \vspace*{6pt}}
\affiliation{Department of Physics, Technion--Israel Institute of
  Technology, \\ 
Technion City, 32000 Haifa, Israel \vspace*{20pt}}
\author{Subhendu Rakshit}\email{srakshit@physics.technion.ac.il}
\affiliation{Department of Physics, Technion--Israel Institute of
  Technology, \\ 
Technion City, 32000 Haifa, Israel \vspace*{20pt}}

\begin{abstract} \vspace*{10pt}
We study neutrino masses and mixing in R-parity violating
supersymmetric models with generic soft supersymmetry breaking
terms. Neutrinos acquire masses from various sources: Tree level
neutrino--neutralino mixing and loop effects proportional to bilinear
and/or trilinear R-parity violating parameters.  Each of these
contributions is controlled by different parameters and have different
suppression or enhancement factors which we identified. Within an
Abelian horizontal symmetry framework these factors are related and
specific predictions can be made. We found that the main contributions
to the neutrino masses are from the tree level and the bilinear loops
and that the observed neutrino data can be accommodated once mild
fine-tuning is allowed.
\end{abstract}

\maketitle

\section{Introduction}
\label{sec:introduction}

Neutrino oscillation experiments indicate that the neutrinos are
massive~\cite{rev}.  The data is best explained with the following set
of parameters~\cite{SKatmo03} 
\beqa \label{fit}
&& \Delta m_{23}^2 = 2.0 \times 10^{-3}~{\rm eV}^2, \qquad \Delta m_{12}^2 =
7.2 \times 10^{-5}~{\rm eV}^2, \\ 
&& \sin^2\theta_{23}=0.5, \qquad  \sin^2\theta_{12}=0.3, \qquad
 \sin^2\theta_{13}<0.074,\nonumber 
\eeqa
where $\Delta m_{ij}^2 \equiv m_i^2-m_j^2$ and $\theta_{ij}$ are the
leptonic mixing angles.  Eq.~(\ref{fit}) tells us that the neutrino
masses exhibit a mild hierarchy and that there is one somewhat small
mixing angle ($\theta_{13}$) and two large mixing angles
($\theta_{12}$ and $\theta_{23}$).

Any theory beyond the Standard Model (SM) needs to explain this
neutrino mass structure.  One of the challenges is to generate large
mixing angles with hierarchical masses.  Generally, small mixing
angles are associated with mass hierarchies and vice versa. This
situation is avoided when the determinant of the mass matrix is much
smaller then its natural value, namely, when there are cancellations
between different terms in the determinant. Such cancellations can
arise naturally in models where different neutrinos acquire masses
from different sources. One such a framework is R-parity Violating
(RPV) supersymmetry~\cite{Fayet:1974pd}, where generically a single
neutrino acquires a mass at the tree level via mixing with the
neutralinos while the other two neutrinos become massive by one-loop
effects.

Neutrino masses in the framework of RPV supersymmetry have been widely
studied~\cite{all}. In the earlier works, the only loop contributions
that were considered are from the loops that depend on trilinear RPV
couplings. Later, it was realized that the effect of
sneutrino--antisneutrino mixing~\cite{ghprl,HKK} can be very important
since it is related to loops that contribute to the neutrino masses
and depend on bilinear RPV parameters~\cite{ghrpv,ghtalk}.  In
Ref.~\cite{Dav-Los} many more loop contributions besides the
``traditional'' trilinear ones were identified. These loops were also
studied in~\cite{numerical,Davidson:1999mc,biloop,Chun}.

In generic RPV models there are too many free parameters and no
specific predictions for the neutrino spectrum can be made. In general
it is even not possible to identify the important contributions to the
neutrino masses.  In this paper we discuss the various contributions
to neutrino masses and identify different suppression and enhancement
factors in each of them. We also study one specific framework, that of
Abelian horizontal symmetry, where specific predictions can be
made. We found that the main contributions to the neutrino masses are
from the tree level and the sneutrino--neutralino loops and that the
model can accommodate the observed data once mild fine-tuning is
allowed.

\section{The model}
\label{sec:the-model}

We start by describing the RPV framework. We follow here the notation
of~\cite{ghrpv} where the model is described in more details.

In order to avoid the bounds from proton stability, we consider the
most general low-energy supersymmetric model consisting of the MSSM
fields that conserves a ${\bf Z_{\,3}}$ baryon triality
\cite{Ibanez:1991pr}.  Such a theory possesses RPV-interactions that
violate lepton number. Once R-parity is violated, there is no
conserved quantum number that distinguishes between the lepton
supermultiplets $\hat L_m$ ($m=1,2,3$) and the down-type Higgs
supermultiplet $\hat H_D$.  It is therefore convenient to denote the
four supermultiplets by one symbol $\hat L_\alpha$ ($\alpha=0,1,2,3$),
with $\hat L_0\equiv \hat H_D$. We use Greek indices to indicate the
four dimensional extended lepton flavor space, and Latin ones for the
usual three dimensional flavor spaces.

The most general renormalizable superpotential is given by:
\beq \label{rpvsuppot}
W=\epsilon_{ij} \left[
-\mu_\alpha \hat L_\alpha^i \hat H_U^j + 
\half\lam_{\alpha\beta m}\hat L_\alpha^i \hat L_\beta^j \hat E_m +
\lp_{\alpha nm} \hat L_\alpha^i \hat Q_n^j  \hat D_m
-h_{nm}\hat H_U^i \hat Q^j_n \hat U_m
\right]\,,
\eeq
where $\hat H_U$ is the up-type Higgs supermultiplet, the $\hat Q_n$
are doublet quark supermultiplets, $\hat U_m$ [$\hat D_m$] are singlet
up-type [down-type] quark supermultiplets and $\hat E_m$ are the
singlet charged lepton supermultiplets.  The coefficients
$\lambda_{\alpha\beta m}$ are antisymmetric under the interchange of
the indices $\alpha$ and $\beta$. Note that the $\mu$-term of the MSSM
[which corresponds to $\mu_0$ in \eq{rpvsuppot}] is now extended to a
four-component vector, $\mu_\alpha$, and that the Yukawa matrices of
the MSSM [which correspond to $\lp_{0ij}$ and $\lam_{0ij}$ in
\eq{rpvsuppot}] are now extended into rank three tensors, $\lp_{\alpha
nm}$ and $\lam_{\alpha\beta m}$.

Next we consider the most general set of renormalizable soft
supersymmetry breaking terms.  In addition to the usual soft
supersymmetry breaking terms of the R-parity Conserving (RPC) MSSM,
one must also add new $A$ and $B$ terms corresponding to the RPV terms
of the superpotential.  In addition, new RPV scalar squared-mass terms
also exist.  As above, we extend the definitions of the RPC terms to
allow indices of type $\alpha$. Explicitly, the relevant terms are
\beq \label{softsusy}
 V_{\rm soft}  = 
 (M^2_{\widetilde L})_{\alpha\beta}\,
          \widetilde L^{i*}_\alpha\widetilde L^i_\beta
  -(\epsilon_{ij} B_\alpha\tilde L_\alpha^i H_U^j +{\rm h.c.}) \nonumber\\
 +  \epsilon_{ij} \bigl[\half A_{\alpha\beta m} \widetilde L^i_\alpha
       \widetilde L^j_\beta \widetilde E_m + A'_{\alpha nm} \widetilde L^i_\alpha\widetilde Q^j_n\widetilde D_m +{\rm
h.c.}\bigl],
\eeq
and we do not present the terms that are unchanged from the RPC (that
can be found, for example, in~\cite{ghrpv}).  Note that the single $B$
term of the MSSM is extended to a four-component vector, $B_\alpha$,
and that the single squared-mass term for the down-type Higgs boson
and the $3\times 3$ lepton scalar squared-mass matrix are now part of
a $4\times 4$ matrix, $(M^2_{\widetilde L})_{\alpha\beta}$.  We
further define
\beq \label{mudef}
|\mu|^2\equiv \sum_\alpha|\mu_\alpha|^2,\qquad
\langle H_U\rangle\equiv \sqhalf v_u, \qquad 
\langle\snu_\alpha\rangle\equiv \sqhalf v_\alpha, \qquad
v_d\equiv |v_\alpha|,
\eeq
with 
\beq \label{vevdef}
v\equiv (|v_u|^2+|v_d|^2)^{1/2}={2m_W\over g}=246~{\rm GeV}\,, \qquad
\tan\beta \equiv {v_u \over v_d}.
\eeq
These vacuum expectation values are determined via the minimum
equations~\cite{ghrpv}.

{}From now on we will work in a specific basis in the space spanned by
$\hat L_\alpha$ such that $v_m=0$ and $v_0=v_d$.  The down-type quark
and lepton mass matrices in this basis arise from the Yukawa couplings
to $\hat H_D$, namely,
\beq \label{ellmassdef}
(m_d)_{nm} = \sqhalf v_d \lp_{0nm}\,, \qquad
(m_\ell)_{nm} = \sqhalf v_d \lam_{0nm}\,.
\eeq
Note that due to the small RPV admixture with the charged Higgsinos,
$(m_\ell)_{nm}$ is not precisely the charged lepton mass matrix
\cite{Davidson:1999mc}. This small effect is not important for our analysis.

In the literature one often finds other basis choices.  The most
common is the one where $\mu_0=\mu$ and $\mu_m=0$.  Of course, the
results for physical observables are independent of the basis
choice. (For basis independent parameterizations of R-parity violation
see~\cite{sachaellis,Ferr,Dav-Los,ghbasis}.)

\section{Neutrino masses}
\label{sec:neutrino-masses}

The neutrino mass matrix receives contributions both at the tree level
and from loops. In the following we review the various kinds of
contributions and identify the leading factors that govern their
magnitudes.

\subsection{Tree level ($\mu\mu$) masses}

\begin{figure}[tb]
\unitlength1mm
\SetScale{2.8}
\begin{boldmath}
\begin{center}
\begin{picture}(60,10)(0,0)
\Line(0,0)(60,0)
\Text(-2,0)[r]{$\nu_i$}
\Text(62,-0.4)[l]{$\nu_j$}
\Text(30,-5)[c]{$\chi_{\alpha}$}
\Text(30,5)[c]{$m_{\chi_{\alpha}}$}
\Text(30,0)[c]{$\times$}
\Text(15,0)[c]{$\bullet$}
\Text(45,0)[c]{$\bullet$}
\Text(15,5)[c]{$\mu_i$}
\Text(45,5)[c]{$\mu_j$}
\end{picture}
\end{center}
\end{boldmath}
\caption{Tree level neutrino mass in the mass
insertion approximation. A blob represents mixing between the neutrino
and the up-type Higgsino. The cross on the neutralino propagator
signifies a Majorana mass term for the neutralino.}
\protect\label{fig:tree-mass}
\end{figure}
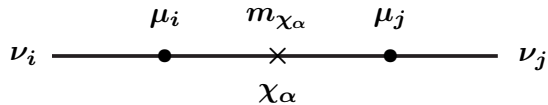

At tree level the neutrino mass matrix receives contributions from RPV
mixing between the neutrinos and the neutralinos, see
Fig.~\ref{fig:tree-mass}. The masses are calculated from the neutral
fermion (neutralinos and neutrinos) mass matrix.  We work
perturbatively, and thus at leading order we do not distinguish
between $\mu$ and $\mu_0$. Then, the tree level neutral fermion mass
matrix, with rows and columns corresponding to $\{\widetilde
B,\widetilde W^3,\widetilde H_U, \nu_\beta \}$, is given by
\cite{bgnn,enrico,ghrpv}:
\beq
\pmatrix{
M_1&0&m_Z\sw v_u/v&-m_Z\sw v_d/v& 0 & 0 & 0\cr
0&M_2&-m_Z\cw v_u/v&m_Z\cw v_d/v& 0 & 0 & 0\cr
m_Z\sw v_u/v&-m_Z\cw v_u/v&0&\mu&~\mu_1  & ~\mu_2 & ~\mu_3\cr
-m_Z\sw v_d/v&m_Z\cw v_d/v&\mu&0 & 0 & 0& 0\cr
0 & 0 &\mu_1&0 & 0 & 0& 0\cr
0 & 0 &\mu_2&0 & 0 & 0& 0\cr
0 & 0 &\mu_3&0 & 0 & 0& 0}\,
\eeq
where $M_1$ is the Bino mass, $M_2$ is the Wino mass,
$\cw\equiv\cos\theta_W$ and $\sw\equiv\sin\theta_W$.  Integrating out
the four neutralinos we get the neutrino mass matrix
\beq
[m_\nu]_{ij}^{(\mu\mu)} = X_T \mu_i \mu_j\,,
\eeq
where 
\beq
X_T = {m_Z^2 m_{\tilde \gamma}\cos^2\beta \over 
\mu(m_Z^2 m_{\tilde \gamma}\sin 2\beta-M_1 M_2 \mu)} 
\sim {\cos^2\beta\over \tilde m}\,,
\eeq
such that $m_{\tilde \gamma}\equiv \cw^2 M_1 + \sw^2 M_2$ and in the
last step we assume that all the relevant masses are at the
electroweak (or supersymmetry breaking) scale, $\tilde m$.  The tree
level neutrino masses are the eigenvalues of $[m_\nu]_{ij}^{(\mu\mu)}$
\beq
m_3^{(T)} = X_T (\mu_1^2 + \mu_2^2+ \mu_3^2)\,, \qquad
m_1^{(T)} = m_2^{(T)}=0\,.
\eeq
Here, and in what follows, we use $m_3\ge m_2
\ge m_1$.

We see that at the tree level only one neutrino is massive. Its mass
is proportional to the RPV parameter $\sum\mu_i^2$ and to
$\cos^2\beta$. For large $\tan \beta$ the latter is a suppression
factor. As we discuss later, this suppression factor can be important.

\subsection{Trilinear ($\lp\lp$ and $\lam\lam$) loops}

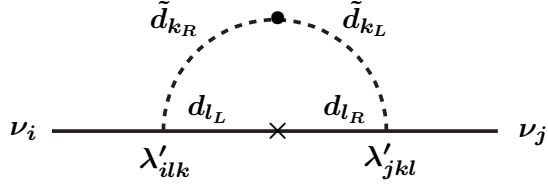
\begin{figure}[tbp]
\unitlength1mm
\SetScale{2.8}
\begin{boldmath}
\begin{center}
\begin{picture}(60,20)(0,0)
\Line(0,0)(60,0)
\Text(15,-4)[c]{$\lambda'_{ilk}$}
\Text(45,-4)[c]{$\lambda'_{jkl}$}
\DashCArc(30,0)(15,0,180){1}
\Text(13,15)[l]{$\tilde{d}_{k_{R}}$}
\Text(18,3)[l]{$d_{l_{L}}$}
\Text(45,15)[r]{$\tilde{d}_{k_{L}}$}
\Text(42,3)[r]{$d_{l_{R}}$}
\Text(-2,0)[r]{$\nu_i$}
\Text(62,-0.4)[l]{$\nu_j$}
\Text(30,0)[c]{$\times$}
\Text(30,15)[c]{$\bullet$}
\end{picture}
\end{center}
\end{boldmath}
\caption[a]{Trilinear loop contribution to the neutrino mass matrix. 
The blob on the scalar line indicates mixing between the left-handed
and the right-handed squarks. A mass insertion on the internal quark
propagator is denoted by the cross.}
\label{trilinear}
\end{figure}

The neutrino mass matrix receives contributions from loops that are
proportional to trilinear RPV couplings, see Fig.~\ref{trilinear}.
These kinds of loops received much attention in the literature.  Here
we only present approximated expressions which are sufficient for our
study. Full results can be found, for example, in
\cite{ghrpv}.

Neglecting quark flavor mixing, the contribution of the $\lp\lp$ loops
is proportional to the internal fermion mass and to the mixing between
left and right sfermions. Explicitly,
\beq
[m_\nu]_{ij}^{(\lp\lp)} \approx \sum_{l,k}\frac{3}{8\pi^2} 
\lambda'_{ilk} \, \lambda'_{jkl} \, 
\frac{m_{d_l}\, \Delta m_{\tilde d_k}^2}{m_{\tilde d_k}^2} 
\sim
\sum_{l,k} \frac{3}{8\pi^2} \lambda'_{ilk} \lambda'_{jkl}
\frac{m_{d_l} m_{d_k}}{\tilde m},
\eeq 
where $m_{\tilde d_k}$ is the average $k$th sfermion mass, $\Delta
m_{{\tilde d}_k}^2$ is the squared mass splitting between the two
$k$th sfermions, and in the last step we used $\Delta m_{{\tilde
d}_k}^2\approx m_{d_k} \tilde{m}$ and $m_{\tilde d_k}\sim\tilde{m}$.
There are similar contributions from loops with intermediate charged
leptons where $\lambda'$ is replaced by $\lambda$ and there is no
color factor in the numerator.

We see that the trilinear loop-generated masses are suppressed by the
RPV couplings $\lambda^{\prime2}$ [$\lam^2$], by a loop factor, and by
two down-type quark [charged lepton] masses. The latter factor, which
is absent in other types of loops, make the trilinear contribution
irrelevant in most cases.

\subsection{Bilinear ($BB$) loop induced masses}

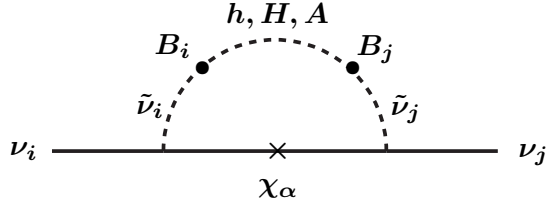
\begin{figure}[tb]
\unitlength1mm
\SetScale{2.8}
\begin{boldmath}
\begin{center}
\begin{picture}(60,20)(0,0)
\Line(0,0)(15,0)
\Line(45,0)(15,0)
\Line(60,0)(45,0)
\DashCArc(30,0)(15,0,180){1}
\Text(13,6)[c]{$\tilde\nu_i$}
\Text(20,11)[c]{$\bullet$}
\Text(16,14)[c]{$B_i$}
\Text(47,5.6)[c]{$\tilde\nu_j$}
\Text(40,11)[c]{$\bullet$}
\Text(43,13.6)[c]{$B_j$}
\Text(-2,0)[r]{$\nu_i$}
\Text(62,-0.4)[l]{$\nu_j$}
\Text(30,18)[c]{$h,H,A$}
\Text(30,0)[c]{$\times$}
\Text(30,-5)[c]{$\chi_{\alpha}$}
\end{picture}
\end{center}
\end{boldmath}
\caption[a]{The $BB$ loop-generated neutrino mass.
Here the blobs denote mixing of the sneutrinos with the neutral Higgs
bosons. The cross on the internal neutralino propagator denotes a
Majorana mass for the neutralino.}
\label{fig-BBloop}
\end{figure}

Neutrinos acquire masses from loops that are proportional to bilinear
RPV couplings. Here we discuss the contributions that are proportional
to two insertions of RPV $B_i$ parameters, see Fig. \ref{fig-BBloop}.
We also refer to the masses induced by these diagrams as the sneutrino
splitting induced masses. The reason is that the two $B$ insertions in
the scalar line also generate splitting between the two sneutrino mass
eigenstates. The contribution of the $BB$ loop diagram is related to
this sneutrino mass splitting~\cite{ghprl,Dav-Los}. In particular, if
the sneutrino splitting vanishes the neutrino is massless.

The one-loop contribution to the neutrino mass matrix from the $BB$
loop is given by
\cite{Dav-Los}
\bea \label{BBloop}
[m_\nu]^{(BB)}_{ij}
& = & \sum_{{\alpha},i,j} 
\frac{g^2 B_i B_j}{4 \cos^2\beta}   
(Z_{{\alpha}2} - Z_{{\alpha}1} g'/g)^2  m_{\chi_{\alpha}}
 \left\{ I_4(m_h, m_{\tilde{\nu}_i},m_{\tilde{\nu}_j}, 
m_{\chi_{\alpha}}) 
\cos^2(\alpha - \beta) \right. \nonumber \\
 && \left. +  
I_4(m_H, m_{\tilde{\nu}_{i}}, m_{\tilde{\nu}_{j}}, 
m_{\chi_{\alpha}}) \sin^2(\alpha - \beta) 
- I_4(m_A, m_{\tilde{\nu}_{i}},m_{\tilde{\nu}_{j}}, 
m_{\chi_{\alpha}}) \right\}\,,
 \label{GH}
\eea
where $Z_{\alpha \beta}$ is the neutralino mixing matrix with
$\alpha,\beta=1,..,4$ and
\beqa \label{Ifour}
I_4(m_{1}, m_{2}, m_{3}, m_4) &=& {1\over m_1^2 -m_2^2} \left[I_3(m_{1},
m_{3}, m_4)-I_3(m_{2}, m_{3}, m_4)\right],\nonumber \\ 
I_3(m_{1}, m_{2}, m_{3}) &=& {1\over m_1^2 -m_2^2} 
\left[I_2(m_{1}, m_{3}) - I_2(m_{2}, m_{3})\right],\nonumber \\ 
I_2(m_{1}, m_{2 }) &=& - {1\over 16\pi^2} {m_1^2\over
m_1^2 -m_2^2} \ln\frac{m_1^2}{m_2^2}.
\eeqa 
Assuming that all the masses in the RHS of Eq. (\ref{BBloop}) are of
the order of the weak scale, we estimate
\beq \label{BBnai}
[m_\nu]_{ij}^{(BB)} \sim \frac{g^2}{64\pi^2 \cos^2\beta} 
{B_i B_j \over  \tilde{m}^3}.
\eeq
In the above estimation, no cancellation between the different Higgs
loops were assumed.  We do, however, expect to have some degree of
cancellation between these loops. To see it note that if the three
$I_4$ functions in (\ref{BBloop}) were equal, $[m_\nu]^{(BB)}_{ij}$
would vanish. The remnant of this effect is a partial cancellation
that becomes stronger in the decoupling limit. Then $\cos^2(\alpha -
\beta)\to 0$ and $m_H\to m_A$ and from \eq{BBloop} we see that
$[m_\nu]^{(BB)}_{ij} \to 0$. We discuss this cancellation in 
Appendix \ref{App-H}.

Next we study the $BB$ loop effect on the neutrino masses.  For this
we rewrite (\ref{BBloop}) as
\beq
[m_\nu]^{(BB)}_{ij}=C_{ij}B_i B_j.
\eeq
If all the elements of the matrix $C_{ij}$ were identical,
$[m_\nu]^{(BB)}_{ij}$ would have only one non-vanishing eigenvalue.
Using (\ref{BBloop}) we see that such a situation arises when the
sneutrinos are degenerate. More generally, we conclude that the
contribution to the light neutrinos receive potentially additional
suppression by a factor proportional to the non-degeneracy in the
sneutrino sector.

In general we expect that $B_\alpha$ is not proportional to
$\mu_{\alpha}$. Then, one neutrino mass eigenstate acquires mass at
tree level, and the other two from bilinear loops where the mass of
the lightest neutrino is proportional to the amount of non-degeneracy
of the sneutrinos. We elaborate more on this effect in Appendix
\ref{App-Deg}.

We conclude that the $BB$ loop-generated masses are suppressed by the
RPV couplings $BB$, by a loop factor and by a possible effect due to
the cancellation between the three Higgs loops.  For large $\tan\beta$
the $BB$ loop is enhanced by $\tan^2\beta$. The third neutrino mass
may get an extra suppression proportional to the non-degeneracy among
the sneutrinos.

\subsection{$\mu B$ loops}

\begin{figure}[tb]
\unitlength1mm
\SetScale{2.8}
\begin{boldmath}
\begin{center}
\begin{picture}(60,20)(0,0)
\Line(0,0)(15,0)
\Line(45,0)(15,0)
\Line(60,0)(45,0)
\DashCArc(30,0)(15,0,180){1}
\Text(8,0)[c]{$\bullet$}
\Text(8,3)[c]{$\mu_i$}
\Text(47,5.6)[c]{$\tilde\nu_j$}
\Text(40,11)[c]{$\bullet$}
\Text(43,14)[c]{$B_j$}
\Text(-2,0)[r]{$\nu_i$}
\Text(62,-0.4)[l]{$\nu_j$}
\Text(25,17)[c]{$h,H,A$}
\Text(30,0)[c]{$\times$}
\Text(12,-5)[c]{$\chi_\alpha$}
\Text(30,-5)[c]{$\chi_{_\beta}$}
\end{picture}
\end{center}
\end{boldmath}
\caption[a]{
Neutrino Majorana mass generated by $\mu B$ loop. The blob on the
external fermion line signifies mixing between a neutrino and a
neutralino. The blob on the internal scalar line stands for mixing
between the sneutrinos and the neutral Higgs bosons. The cross on the
internal neutralino line denotes, as before, a Majorana mass term for
the neutralino. There exists other diagrams with $i\leftrightarrow
j$.}
\label{fig-muBloop}
\end{figure}
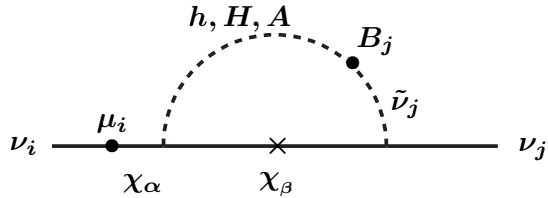

Another type of diagrams which induce neutrino masses from mixing
between the sneutrinos and the neutral Higgs bosons
is given in Fig. \ref{fig-muBloop}.
The contribution from this diagram to the neutrino mass matrix is
given by~\cite{Dav-Los}\footnote{Note that we disagree with
\cite{Dav-Los} on the sign of the term that originate from the $A$
loop.}
\bea
[m_\nu]_{ij}^{(\mu B)}&=& \sum _{\alpha, \beta}
 {{g^2} \over  4 \cos \beta} 
 \mu_{i}  B_j 
{m_{\chi_{_\beta}} \over m_{\chi_{\alpha}}} 
Z_{\alpha 3} ( Z_{\beta 2} -  Z_{\beta 1} g'/g)   
\nonumber \\ &&
\bigg\{-\Big[ 
 Z_{\alpha 4}( Z_{\beta 2}-  Z_{\beta 1} g'/g) \sin \alpha +
   ( Z_{\alpha 2}-  Z_{\alpha 1} g'/g)Z_{\beta 3} \cos \alpha
\nonumber \\&& \quad\quad
+ ( Z_{\alpha 2}-  Z_{\alpha 1} g'/g)  Z_{\beta 4}\sin \alpha \Big]
  \cos(\alpha - \beta) \, 
I_3(m_h,m_{\chi_{_\beta}}, m_{\tilde{\nu}_j})
\nonumber  \\ &&\quad
+ \Big[  Z_{\alpha 4}( Z_{\beta 2}-  Z_{\beta 1} g'/g) \cos \alpha
 -( Z_{\alpha 2}-  Z_{\alpha 1} g'/g)  Z_{\beta 3}\sin \alpha
\nonumber \\&& \quad\quad
+( Z_{\alpha 2}-  Z_{\alpha 1} g'/g) Z_{\beta 4} \cos \alpha \Big]
   \sin(\alpha - \beta) \,
I_3(m_H,m_{\chi_{_\beta}}, m_{\tilde{\nu}_j})
\nonumber \\ && \quad
+ \Big[ Z_{\alpha 4}( Z_{\beta 2}-  Z_{\beta 1} g'/g) \sin \beta
 + ( Z_{\alpha 2}-  Z_{\alpha 1} g'/g)Z_{\beta 3} \cos \beta
\nonumber \\&& \quad\quad
+ ( Z_{\alpha 2}-  Z_{\alpha 1} g'/g)  Z_{\beta 4}\sin \beta \Big] \,
I_3(m_A,m_{\chi_{_\beta}}, m_{\tilde{\nu}_j}) 
\bigg\}  + (i \leftrightarrow j)
\label{muB-term}
\eea
Assuming that all the masses are at the weak scale, this contribution 
to the neutrino mass matrix is given approximately by~\cite{Dav-Los}
\beq
[m_\nu]_{ij}^{(\mu B)}\sim \frac{g^2}{64\pi^2 \cos\beta}\, 
\frac{\mu_i B_j+\mu_j B_i}{\tilde m^2}.
\eeq
In the flavor basis these diagrams are expected to yield similar
contributions to the $BB$ loops.  Yet, as pointed out in~\cite{Chun},
due to the dependence on $\mu_i$, the $\mu B$ loop contribution to
the neutrino masses is sub-leading. See Appendix \ref{App-mu} for
details.

Similar to the $BB$ loops, also in the $\mu B$ loop there is a partial
cancellation between the different Higgs loops. In the decoupling
limit this can be seen from Eq. (\ref{muB-term}).  Since, as we just
mentioned, the effect of the $\mu B$ loop is sub-leading, we do not
elaborate on the decoupling effect.

We conclude that the $\mu B$ loop-generated masses are suppressed by
the RPV couplings $\mu B$, by a loop factor and by a possible effect
due to Higgs decoupling.  In the case where the tree level
contribution is dominant, their effect on the neutrino masses is
second order in these small parameters.

\subsection{Other loops}

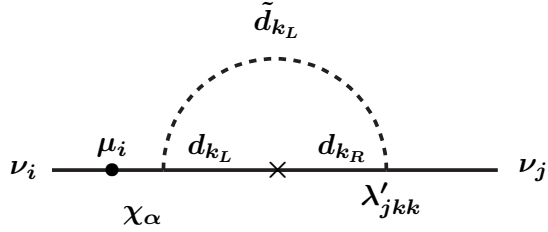
\begin{figure}[tb]
\unitlength1mm
\SetScale{2.8}
\begin{boldmath}
\begin{center}
\begin{picture}(60,20)(0,0)
\Line(0,0)(15,0)
\Text(8,0)[c]{$\bullet$}
\Text(8,3)[c]{$\mu_i$}
\Line(45,0)(15,0)
\Line(60,0)(45,0)
\Text(45,-4)[c]{$\lambda'_{jkk}$}
\DashCArc(30,0)(15,0,180){1}
\Text(18,3)[l]{$d_{k_{L}}$}
\Text(30,20)[c]{$\tilde{d}_{k_{L}}$}
\Text(42,3)[r]{$d_{k_{R}}$}
\Text(-2,0)[r]{$\nu_i$}
\Text(62,0)[l]{$\nu_j$}
\Text(30,0)[c]{$\times$}
\Text(12,-6)[c]{$\chi_{\alpha}$}
\end{picture}
\end{center}
\end{boldmath}
\caption[b]{
Neutrino Majorana mass generated by $\mu \lp$ loop. The blob on the
external fermion line signifies a mixing between a neutrino and a
up-type Higgsino which is then converted to a gaugino. The cross on
the internal fermion line stands for a Dirac mass insertion. There
exists another diagram with $i\leftrightarrow j$.}
\label{fig-mulam}
\end{figure}

There are many other loops that contribute to the neutrino masses
\cite{Dav-Los}. Almost all of them are suppressed by at least two
Yukawa interactions, and are therefore likely to be negligible. 

There is only one contribution to the neutrino masses that depends on both
bilinear and trilinear couplings and is suppressed by only one Yukawa
coupling. This diagram is shown in Fig.~\ref{fig-mulam}.
Neglecting squark flavor mixing, the $\mu\lp$ contribution to the
neutrino mass matrix is~\cite{Dav-Los}
\beq
[m_\nu]_{ij}^{(\mu\lp)} \approx \sum_k \frac{3}{16 \pi^2}\, g\, m_{d_k}\,
 \frac{\mu_i \lambda'_{jkk}+\mu_j \lambda'_{ikk}}   {\tilde m}.
\eeq
There are similar contributions from diagrams with $\lam$ instead of
$\lp$ couplings, where leptons and sleptons are running in the loop. 

We see that the $\mu \lambda'$ diagrams are suppressed by the RPV
couplings $\mu \lambda'$, by a loop factor and by one Yukawa
coupling.  Also here, similar to the case of the $\mu B$ loops,
once the tree level effect
is taken into account, 
the $\mu \lp$ and $\mu \lam$ contributions to the 
light mass eigenstates are second order in the above mentioned
suppression factors.

\subsection{Model independent considerations}
As we discussed, there are many contributions to the neutrino masses
that are suppressed by different small parameters. In general, the
leading effects are model dependent. Nevertheless, here we make some
general remarks.

One neutrino is massive at the tree level and unless $\tan\beta$ is very
large, this is the dominant contribution to $m_3$.  The other
neutrinos get masses at the loop level. Despite the partial cancellation
between different Higgs loops, we expect the $BB$ loops to be the
dominant one. All the other contributions are generically suppressed
compared to it due to the following reasons:
\begin{itemize}
\item
The $\mu B$ loop contribution to the light neutrino masses is second order
in the small ratio between the loop-induced mass and the tree level
one.
\item
The $\lp\lp$ and $\lam\lam$ diagrams are doubly
Yukawa suppressed. 
\item
The $\mu \lp$ and $\mu \lam$ diagrams are singly Yukawa suppressed, and
similar to the situation with the $\mu B$ loops, their contributions to
the light neutrino masses are second order in the suppression factor.
\end{itemize}
Therefore, while there are several caveats as explained above, the
situation is likely to be as follows: The heaviest neutrino mass,
$m_3$, arises at the tree level. The major contribution to $m_2$ is
from the $BB$ loops. For non-degenerate sneutrinos, $m_1$ is also
generated by the $BB$ loops. For degenerate sneutrinos, however, $m_1$
is very small and the major contribution to it can arise from any of
the other sources. Note that since neutrino oscillation data are not
sensitive to the lightest neutrino mass, our ignorance of the
mechanism that generate $m_1$ is not problematic.

In the following we consider a specific model where we can explicitly
check the relevance of the different contributions.

\section{Horizontal symmetry}
\label{sec:horizontal-symmetry}

We work in the Abelian horizontal symmetry framework~\cite{LNS}. The
horizontal symmetry, $H$, is explicitly broken by a small parameter
$\lambda$ to which we attribute charge $-1$. This can be viewed as the
effective low energy theory that comes from a supersymmetric extension
of the Froggatt--Nielsen mechanism at a high scale~\cite{FrNi}. Then,
the following selection rules apply: ($a$) Terms in the superpotential
that carry charge $n\geq0$ under $H$ are suppressed by $O(\lam^n)$,
while those with $n<0$ are forbidden by holomorphy; ($b$)
Soft supersymmetry breaking terms that carry charge $n$ under $H$ are
suppressed by $O(\lam^{|n|})$.  For simplicity, in the following we
assume that the horizontal charges of all the MSSM superfields are
non-negative.

We identify the down-type Higgs doublet with the doublet superfield
that carries the smallest charge, which we choose to be
$L_0\equiv H_d$. (To simplify the notation what we denote before as
$\hat L_0$ is now $L_0$.)  We order the remaining doublets according
to their charges:
\beq \label{orderH}
H(L_1)\geq H(L_2)\geq H(L_3)\geq H(H_d)\geq0\,.
\eeq
Similar ordering is made for the three generations of the charged
leptons and quarks.

Our methods of analyzing lepton and neutralino mass matrices are
described in detail in~\cite{bgnn} and~\cite{lmmm}, respectively.
Specifically, we use the above mentioned selection rules to estimate
the magnitude of the relevant parameters
\beqa \label{selectm}
\mu_\alpha &\sim& \tilde\mu\lam^{H(L_\alpha)+H(H_u)} \,, \\
B_\alpha&\sim&\tilde m^2\lam^{H(L_\alpha)+H(H_u)}\,, \nonumber \\
(M^2_{\widetilde L})_{\alpha\beta} &\sim&
    \tilde m^2\lam^{|H(L_\beta)-H(L_\alpha)|}\,,\nonumber \\
\lp_{\alpha jk}&\sim&
\lam^{H(L_\alpha)+H(Q_j)+H(\bar d_k)}\,,\nonumber \\
\lam_{\alpha\beta k}&\sim&
\lam^{H(L_\alpha)+H(L_\beta)+H(\bar\ell_k)}\,.\nonumber 
\eeqa
Here $\tilde\mu$ is the natural scale for the $\mu$ terms. We assume
that $\tilde\mu=O(\tilde m)$ and since $\mu_0$ is phenomenologically
required to be also of $O(\tilde m)$, we take $H(H_d)=H(H_u)=0$.
Then we get
\beq
{\mu_i \over \mu_0} \sim {B_i \over B_0} \sim {(M^2_{\widetilde L})_{0i}
\over (M^2_{\widetilde L})_{\alpha \alpha}} \sim
{v\, \lam_{ijk} \over (m_\ell)_{jk}} \sim
{v \, \lp_{ijk} \over (m_d)_{jk}} \sim \lam^{H(L_i)}\,.
\eeq
We see that all the RPV parameters are suppressed by a common factor
compared to their corresponding RPC parameters. In particular, this
implies that the RPV trilinear couplings are very small since they
are related to the small RPC Yukawa couplings.

Several other parameters are expected to be of $O(1)$ in the
horizontal symmetry framework: In particular,
\beq
\cos\beta, \qquad \epsilon_H, \qquad \epsilon_D,
\eeq
such that $\epsilon_H$ is the suppression due to the effect of Higgs
decoupling (defined in \eq{def-eps-H}) and $\epsilon_D$ is the
suppression due to sneutrino degeneracy (defined in \eq{def-eps-D}).
Yet, in the following we keep them in order to understand what
parameters are needed to be fine-tuned in order to get a viable model.

Now we can estimate the order of magnitude of the different
contributions to the neutrino mass matrix 
\bea \label{estra}
[m_\nu]_{ij}^{(\mu\mu)} &\sim& m_0\,\cos^2\beta\, \lam^{H(L_i) + H(L_j)}
\,, \\[5pt] 
[m_\nu]_{ij}^{(BB)} &\sim& {m_0\over \cos^2\beta}\,\epsilon_L \epsilon_H
\lam^{H(L_i) + H(L_j)}\,,
\nonumber\\[5pt]
[m_\nu]_{ij}^{(\mu B)} &\sim& {m_0\over\cos\beta} \,\epsilon_L \epsilon_H
 \lam^{H(L_i) + H(L_j)} \,, \nonumber\\[5pt]
[m_\nu]_{ij}^{(\lp\lp)} &\sim& m_0\,\epsilon_L  \left({m_b \over
v}\right)^4 \lam^{H(L_i) + H(L_j)}  \,, \nonumber\\[5pt]
[m_\nu]_{ij}^{(\mu\lp)} &\sim& m_0\,\epsilon_L \left({m_b \over
v}\right)^2 \lam^{H(L_i) + H(L_j)}  \,, \nonumber
\eea
where $\epsilon_L\sim 10^{-2}$ is the loop suppression factor. While
$\epsilon_H$ and $\epsilon_L$ are in general different for the
different contributions, we expect them to be of the same order and thus we
omit their identification indices. In our simple model the overall
scale $m_0 \sim O(\tilde m)$. It can be much smaller if there is a
mechanism that generates a common suppression factor for all the RPV
parameters.

In order to get the relative importance of the different contributions 
to the neutrino masses we note the following points:
\begin{itemize}
\item
There is a common factor,  $m_0 \,\lam^{H(L_i) + H(L_j)}$, to
all the contributions. 
\item
Within the horizontal symmetry framework $[m_\nu]_{ij}^{(BB)} \sim
[m_\nu]_{ij}^{(\mu B)}$.  As explained above, as long as the tree
contribution is dominant, this implies that the effect of
$[m_\nu]_{ij}^{(\mu B)}$ on the neutrino masses is negligible.
\item
Due to the extra Yukawa suppressions $[m_\nu]_{ij}^{(\mu\lp)} >
[m_\nu]_{ij}^{(\lp\lp)}$. However, $[m_\nu]_{ij}^{(\mu\lp)}$ contributes
to the neutrino mass at second order in the suppression factors.
Using \eq{estra} we get
the ratio of these two contributions to the lightest neutrino mass
\beq
{m_1^{(\mu\lp)} \over m_1^{(\lp\lp)}} \sim {\epsilon_L  \over  \cos^2\beta}\,.
\eeq
We conclude that unless $\cos\beta$ is very small, the contribution of
the $\lp\lp$ loops to the neutrino masses is more important than that
of the $\mu\lp$ ones.
\end{itemize}
We see that there are three possible important contributions to the
neutrino masses, $[m_\nu]_{ij}^{(\mu\mu)}$, $[m_\nu]_{ij}^{(BB)}$ and
$[m_\nu]_{ij}^{(\lp\lp)}$ . Their relative effects are controlled by
$\cos\beta$, $\epsilon_H$, and $\epsilon_L$, see \eq{estra}.
We assume that $\cos\beta$, $\epsilon_H$ and
$\epsilon_D$ are not very small. Then, the $\lp\lp$ loops can be
neglected and we get the order of magnitude of the neutrino masses as 
\beqa
m_3 &\sim& m_0 \cos^2\beta \,\lam^{2H(L_3)}, \\[5pt]
m_2 &\sim& \frac{m_0}{\cos^2\beta}\,  \epsilon_L \epsilon_H\lam^{2H(L_2)}, 
\nonumber \\[5pt]
m_1 &\sim& \frac{m_0}{\cos^2\beta}\, \epsilon_L \epsilon_H \epsilon_D \lam^{2H(L_1)} . 
\nonumber
\eeqa
If $\epsilon_D$ is very small then the lightest neutrino mass is
dominated by the  $\lp\lp$ loop contribution, 
\beq
m_1 \sim m_0 \lam^{2H(L_1)}
\epsilon_L \left({m_b\over v}\right)^4 \,.
\eeq

Next we check whether the neutrino data can be explained in our model.
The mixing angles are given by~\cite{lmmm} 
\beq
\sin \theta_{ij} \sim  \lam^{|H(L_i)-H(L_j)|}.
\eeq
The requirement that 
$\theta_{23}$ and $\theta_{12}$ are large~\cite{SKatmo03} implies that
\beq
H(L_3)=H(L_2)=H(L_1).
\eeq
A potential problem is that this choice of horizontal charges also predicts
large $\theta_{13}$.
In order to generate a viable neutrino mass spectrum we require that
$m_3 \sim 10^{-1}\;{\rm eV}$ and 
$m_2 \sim 10^{-2}\;{\rm eV}$. This is the case when 
\beq \label{ptoyu}
m_0 \lam^{2H(L_3)} \cos^2\beta \sim 10^{-1}\;{\rm eV}\,,
\eeq
and
\beq \label{ratpp}
{\cos^4\beta \over \epsilon_H} \sim 10^{-1},
\eeq
where we used $\epsilon_L \sim 10^{-2}$.

The requirement in \eq{ptoyu} can be met once appropriate input
parameters (or charges) are chosen.  The ratio in (\ref{ratpp}) as
well as the requirement of small $\theta_{13}$, however, required mild
fine-tuning. Within our model both are predicted to be of order $O(1)$
while the data suggest that they are $O(10^{-1})$.

Here we consider only a simple model base on a $U(1)_H$ symmetry. In
more elaborated models, like those with more complicated symmetry
group, e.g. $U(1)\times U(1)$ or discrete symmetry group~\cite{yyy},
one may be able to achieve a viable model with less fine-tuning.

\section{Conclusions}
\label{sec:conclusions}

RPV supersymmetric models provide an alternative to the see-saw
mechanism. One virtue of RPV models is that they naturally provide a
mechanism for large mixing with hierarchy, as indicated by the data.

We study the magnitudes of various sources of neutrino masses in RPV
models. There are several parameters that determine them and therefore
there are several suppression factors in each of them. Thus, their
relative importance is model dependent. Due to the Yukawa suppression
of the trilinear loops, it is generally likely that the tree level and
the $BB$ loops are the dominant contributions.

We study one specific model with an Abelian horizontal
symmetry. In this model indeed the tree level and the $BB$
one-loop contributions are the dominant ones. We find that such a
model can describe the neutrino data as long as mild fine-tuning is
permitted.

\acknowledgments
We thank Howie Haber, Sourov Roy and Jure Zupan for helpful
discussions. S.R. thanks the Fermilab and SLAC theory groups for
hospitality while parts of this work were completed. He also
acknowledges financial support from the Lady Davis Fellowship Trust. The
work of Y.G. is supported by the Department of Energy, contract
DE-AC03-76SF00515 and by the Department of Energy under grant
No.~DE-FG03-92ER40689.

\appendix

\section{Higgs cancellation in the $BB$ loops}\label{App-H}
Here we study the cancellation between the three $BB$ loops of
Fig~\ref{BBloop}. The weighted sum of the three Higgs propagators,
before integrating over the internal momenta $k$, is
\beq
P_S=\frac{1}{k^2-m_h^2} \cos^2(\alpha-\beta) + \frac{1}{k^2-m_H^2} 
\sin^2(\alpha-\beta) - \frac{1}{k^2-m_A^2}.
\eeq
For simplicity we use the tree level relations~\cite{hunter,gh1}
\beq
\cos^2(\alpha-\beta)=\frac{m_h^2(m_Z^2-m_h^2)}{m_A^2(m_H^2-m_h^2)},\qquad
m_Z^2-m_h^2=m_H^2-m_A^2,
\eeq
and we obtain
\beq \label{PSff}
P_S=\frac{-k^2(m_Z^2-m_h^2)(m_A^2-m_h^2)}
{m_A^2(k^2-m_H^2)(k^2-m_A^2)(k^2-m_h^2)}.
\eeq
Consider the decoupling limit where $m_H\sim m_A \gg m_h \sim m_Z$. In
that limit the $H$ and $A$ propagators scale like one over their heavy
mass squared. {}From Eq. (\ref{PSff}) we see that the weighted sum
scales like one over the heavy mass to the fourth power.

While the partial cancellation is more severe in the decoupling limit,
it also occurs far away from that limit. The reason is
that the $I_4$ function, defined in \eq{Ifour}, is not very sensitive
to variation in one of its arguments as long as it is not the largest
one.

We have checked the effect of the summation over the different Higgs
mediated diagrams numerically. We define the following measure of the
suppression factor
\beq \label{def-eps-H}
\epsilon_H \equiv 
\left|{I(m_h) \cos^2(\alpha - \beta) + I(m_H) \sin^2(\alpha - \beta) - I(m_A)
\over
|I(m_h)| \cos^2(\alpha - \beta) + |I(m_H)| \sin^2(\alpha - \beta) +
|I(m_A)|}\right|\,,
\eeq
where $I(x)\equiv I_4(x, m_{\tilde{\nu}_{i}}, m_{\tilde{\nu}_{j}},
m_{\chi_{\alpha}})$ [see \eq{BBloop}], and the $i,j,\alpha$ indices
of $\epsilon_H$ are implicit.  While the tree level relation is a good
approximation of the effect, in the numerical calculation we use the
two-loop spectrum for the Higgs boson masses and mixing angles
\cite{suspect}. Some representative numbers are presented in
Table~\ref{Hcan-table}. 

We also checked the effect of $\tan\beta$. We found that
$\epsilon_H$ decreases as $\tan\beta$ increases.  Thus, the
sensitivity of $[m_\nu]^{(BB)}_{ij}$ to $\tan\beta$ is reduced. On one
hand it scales like $1/\cos^2\beta$ [see \eq{BBloop}], and on the
other hand the cancellation between the different Higgs loops becomes 
stronger for large $\tan\beta$. In fact, using the tree level Higgs
mass relations, we found that asymptotically $\epsilon_H \propto
\cos^2\beta$. Thus, at the tree level in the
$\tan\beta\to\infty$ limit, $[m_\nu]^{(BB)}_{ij}$ is independent of
$\tan\beta$.

\begin{table}[tbp]
\begin{tabular}{|c|c|c|c|c|}
\hline
$\tan\beta$ & $m_h$ &$m_A$ & $m_H$ & $\epsilon_H\times 10^3$\\
\hline
4& 92 & 184 & 190 & $9.1$ \\
2&  81&  426 &  430 & $4.7$ \\
20&  106&  285&  284& $1.1$ \\
14&  106&  294&  294& $0.3$ \\
\hline
\end{tabular}
\caption{Numerical values of the suppression factor due to 
the cancellation between different Higgs contributions in the $BB$
loops. We used $m_{\snu_1}=100$ GeV, $m_{\snu_2}=200$ GeV and
$m_{\chi_\alpha}= 300$ GeV. 
}
\label{Hcan-table}
\end{table}

\section{The suppression due to sneutrino degeneracy }\label{App-Deg}
Here we study the effect of the sneutrino degeneracy on the light mass
eigenstate from the $BB$ loops. We assume that the heaviest neutrino
acquires large mass at the tree level. Then, for simplicity, 
we deal only with the loop contribution to the first two generations.

We define the mass-squares of the two sneutrinos as 
\beq
(m^2_{\snu})_{1,2} \equiv m_{\snu}^2 (1\pm\Delta).
\eeq
Computing the $BB$ one-loop contributions up to order $\Delta^2$, we
get a mass matrix of the following form:
\beq
f_1 \; 
\pmatrix{
B_1 B_1 & ~B_1 B_2 \cr B_2 B_1 & ~B_2 B_2}
+ \Delta f_2  
\pmatrix{
B_1 B_1 & 0  \cr 0 & -B_2 B_2}
+ \Delta^2  f_3 
\pmatrix{
3B_1 B_1 &  ~B_1 B_2  \cr  B_2 B_1 & ~3B_2
B_2}
+ {\cal O}(\Delta^3)
\eeq
where
\beq
f_1 =m_{Deg}|_{\Delta\rightarrow 0}\;, \qquad
f_2 = \left.\frac{\partial m_{Deg}}{\partial
\Delta}\displaystyle\right|_{\Delta\rightarrow 0},\qquad
f_3 = \left.\frac{1}{2}\,\frac{\partial^2 m_{Deg}}{\partial
\Delta^2}\displaystyle\right|_{\Delta\rightarrow 0}.
\eeq
and
\bea
m_{Deg}(\Delta)& =& 
\sum_{\alpha} g^2 \frac{1}{4\,\cos^2\beta} (Z_{\alpha 2}
- Z_{\alpha 1} g'/g)^2 m_{\chi_{\alpha}}\Big[I_4(m_h,m,
m,m_{\chi_{\alpha}})
\cos^2(\alpha-\beta) \nonumber \\
       &+&I_4(m_H,m,m,m_{\chi_{\alpha}})
\sin^2(\alpha-\beta)  - I_4(m_A,m,m,m_{\chi_{\alpha}}) \Big],
\eea
where $m^2=m_{\snu}^2(1+\Delta)$.
After diagonalization, we get the following masses:
\bea
m_{2} &=& (B_1^2+B_2^2)\, f_1 + {\cal O}(\Delta),\nonumber\\
m_{1}&=&\frac{B_1^2 B_2^2}{B_1^2+B_2^2} (4 f_1 f_3 - f_2^2)
\Delta^2 +  {\cal O}(\Delta^3).
\label{masses}
\eea
We see that the dominant contribution to $m_{2}$ is the same as that
in the degenerate case. The leading contribution to $m_{1}$, on the
other hand, is proportional to the square of the sneutrino mass   
splitting.

We define the following measure of the degeneracy suppression
\beq \label{def-eps-D}
\epsilon_D\equiv {m_1 \over m_2},
\eeq
which is given by
\beq
\epsilon_D = f_c\, 
\frac{B_1^2 B_2^2}{(B_1^2+B_2^2)^2}\, \Delta^2, 
\qquad f_c= {4 f_1 f_3-f_2^2 \over f_1}.
\eeq
We have checked numerically and found that typically $f_c \sim 0.1$.
Thus, in addition to the $\Delta^2$ suppression, the lightest
neutrino mass is also suppressed by $f_c$.

\section{$\mu_i$ dependent one-loop contributions}\label{App-mu}
Here we explain why when the tree level is the dominant contribution
to the neutrino mass matrix, the effect of loops that have one $\mu_i$
insertion are small. They appear only at second order in the ratio
between the loop contribution to the mass matrix and the tree level
one.  This effect was also discussed in
\cite{Chun}.

We consider a two generation case with only one type of one-loop
contribution at a time. First we assume that we have the following
two contributions
\beq
[m_\nu]^{(\mu\mu)}_{ij} = C \mu_i \mu_j, \qquad
[m_\nu]^{(V\mu)}_{ij} = C \varepsilon_L (\mu_i V_j + \mu_j V_i) 
\eeq
where $C$ is a constant and $V$ is a normalized general vector in
flavor space such that $|V|=|\mu|$.  For example, in the case of the
$\mu B$ diagram, $V_i$ corresponds to the product of $B_i$ with the
loop function.  The fact that the tree level is dominant is encoded by
the choice $\varepsilon_L \ll 1$. The mass matrix is then
\beq
m_\nu = C
\pmatrix{
\mu_1^2 + 2 \varepsilon_L V_1 \mu_1 & \mu_1 \mu_2 + \varepsilon_L
(V_1\mu_2 +V_2 \mu_1) \cr \mu_1 \mu_2 + \varepsilon_L
(V_1\mu_2 +V_2 \mu_1) &\mu_2^2 + 2 \varepsilon_L V_2 \mu_2}
\eeq
We see that the ratio of the two mass eigenstates is
\beq\label{doub-ep}
{m_1 \over m_2} \sim O\left(\varepsilon_L^2\right).
\eeq

Next consider a case where the loop effect is generated without any
$\mu_i$ insertion. For example
\beq  \label{dfrp}
[m_\nu]^{(\mu\mu)}_{ij} = C \mu_i \mu_j, \qquad
[m_\nu]^{(VU)}_{ij} = C^{VU}_{ij} \varepsilon_L (V_i U_j+V_j U_i),
\eeq
where $U$ is another normalized vector and $C \sim C^{VU}_{ij}$ for
any $i$ and $j$. For
$m_\nu=[m_\nu]^{(\mu\mu)}+[m_\nu]^{(VU)}$ we generally get
\beq \label{sin-ep}
{m_1 \over m_2} \sim O(\varepsilon_L).
\eeq
Note that the above holds also for $U=V$.

Comparing \eqs{sin-ep}{doub-ep} we see that diagrams with one $\mu_i$
insertion are unlikely to affect the neutrino masses significantly.



\begin{thebibliography}{99}

\bibitem{rev}
For reviews see, for example,
M.~C.~Gonzalez-Garcia and Y.~Nir,
Rev.\ Mod.\ Phys.\  {\bf 75}, 345 (2003)
[hep-ph/0202058];
Y.~Grossman,
hep-ph/0305245;
V.~Barger, D.~Marfatia and K.~Whisnant,
Int.\ J.\ Mod.\ Phys.\ E {\bf 12}, 569 (2003)
[hep-ph/0308123].

\bibitem{SKatmo03}
M.~C.~Gonzalez-Garcia and C.~Pena-Garay,
hep-ph/0306001;
M.~Maltoni, T.~Schwetz, M.~A.~Tortola and J.~W.~Valle,
hep-ph/0309130; 
A.~Bandyopadhyay, S.~Choubey, S.~Goswami, S.~T.~Petcov and D.~P.~Roy,
hep-ph/0309174; 
P.~C.~de Holanda and A.~Y.~Smirnov,
hep-ph/0309299.



\bibitem{Fayet:1974pd}
P.~Fayet,
Nucl.\ Phys.\ B {\bf 90}, 104 (1975);
Phys.\ Lett.\ B {\bf 69}, 489 (1977);
Phys.\ Lett.\ B {\bf 76}, 575 (1978).

\bibitem{all}
C.~S.~Aulakh and R.~N.~Mohapatra,
Phys.\ Lett.\ B {\bf 119}, 136 (1982);
L.~J.~Hall and M.~Suzuki,
Nucl.\ Phys.\ B {\bf 231}, 419 (1984);
I.~H.~Lee,
Phys.\ Lett.\ B {\bf 138}, 121 (1984);
Nucl.\ Phys.\ B {\bf 246}, 120 (1984);
G.~G.~Ross and J.~W.~Valle,
Phys.\ Lett.\ B {\bf 151}, 375 (1985);
J.~R.~Ellis, G.~Gelmini, C.~Jarlskog, G.~G.~Ross and J.~W.~Valle,
Phys.\ Lett.\ B {\bf 150}, 142 (1985); 
S.~Dawson,
Nucl.\ Phys.\ B {\bf 261}, 297 (1985);
A.~Santamaria and J.~W.~Valle,
Phys.\ Lett.\ B {\bf 195}, 423 (1987);
K.~S.~Babu and R.~N.~Mohapatra,
Phys.\ Rev.\ Lett.\  {\bf 64}, 1705 (1990);
R.~Barbieri, M.~M.~Guzzo, A.~Masiero and D.~Tommasini,
Phys.\ Lett.\ B {\bf 252}, 251 (1990); 
E.~Roulet and D.~Tommasini,
Phys.\ Lett.\ B {\bf 256}, 218 (1991);
K.~Enqvist, A.~Masiero and A.~Riotto,
Nucl.\ Phys.\ B {\bf 373}, 95 (1992);
J.~C.~Romao and J.~W.~Valle,
Nucl.\ Phys.\ B {\bf 381}, 87 (1992); 
R.~M.~Godbole, P.~Roy and X.~Tata,
Nucl.\ Phys.\ B {\bf 401}, 67 (1993)
[hep-ph/9209251]; 
A.~S.~Joshipura and M.~Nowakowski,
Phys.\ Rev.\ D {\bf 51}, 2421 (1995)
[hep-ph/9408224]; 
Phys.\ Rev.\ D {\bf 51}, 5271 (1995)
[hep-ph/9403349]; 
M.~Nowakowski and A.~Pilaftsis,
Nucl.\ Phys.\ B {\bf 461}, 19 (1996)
[hep-ph/9508271]; 
F.~M.~Borzumati, Y.~Grossman, E.~Nardi and Y.~Nir,
Phys.\ Lett.\ B {\bf 384}, 123 (1996)
[hep-ph/9606251]; 
M.~Drees, S.~Pakvasa, X.~Tata and T.~ter Veldhuis,
Phys.\ Rev.\ D {\bf 57}, 5335 (1998)
[hep-ph/9712392]; 
R.~Adhikari and G.~Omanovic,
Phys.\ Rev.\ D {\bf 59}, 073003 (1999);
B.~Mukhopadhyaya, S.~Roy and F.~Vissani,
Phys.\ Lett.\ B {\bf 443}, 191 (1998) 
[hep-ph/9808265]; 
K.~Choi, K.~Hwang and E.~J.~Chun,
Phys.\ Rev.\ D {\bf 60}, 031301 (1999)
[hep-ph/9811363]; 
S.~Rakshit, G.~Bhattacharyya and A.~Raychaudhuri,
Phys.\ Rev.\ D {\bf 59}, 091701 (1999)
[hep-ph/9811500]; 
D.~E.~Kaplan and A.~E.~Nelson,
JHEP {\bf 0001}, 033 (2000)
[hep-ph/9901254]; 
A.~S.~Joshipura and S.~K.~Vempati,
Phys.\ Rev.\ D {\bf 60}, 111303 (1999)
[hep-ph/9903435]; 
S.~Y.~Choi, E.~J.~Chun, S.~K.~Kang and J.~S.~Lee,
Phys.\ Rev.\ D {\bf 60}, 075002 (1999)
[hep-ph/9903465]; 
A.~Datta, B.~Mukhopadhyaya and S.~Roy,
Phys.\ Rev.\ D {\bf 61}, 055006 (2000)
[hep-ph/9905549]; 
A.~Abada and M.~Losada,
Nucl.\ Phys.\ B {\bf 585}, 45 (2000)
[hep-ph/9908352]; 
O.~Haug, J.~D.~Vergados, A.~Faessler and S.~Kovalenko,
Nucl.\ Phys.\ B {\bf 565}, 38 (2000)
[hep-ph/9909318]; 
E.~J.~Chun and S.~K.~Kang,
Phys.\ Rev.\ D {\bf 61}, 075012 (2000)
[hep-ph/9909429]; 
F.~Takayama and M.~Yamaguchi,
Phys.\ Lett.\ B {\bf 476}, 116 (2000)
[hep-ph/9910320]; 
R.~Kitano and K.~y.~Oda,
Phys.\ Rev.\ D {\bf 61}, 113001 (2000)
[hep-ph/9911327]; 
M.~Hirsch, M.~A.~Diaz, W.~Porod, J.~C.~Romao and J.~W.~Valle,
Phys.\ Rev.\ D {\bf 62}, 113008 (2000)
[Erratum-ibid.\ D {\bf 65}, 119901 (2002)]
[hep-ph/0004115]; 
J.~M.~Mira, E.~Nardi, D.~A.~Restrepo and J.~W.~Valle,
Phys.\ Lett.\ B {\bf 492}, 81 (2000)
[hep-ph/0007266];
T.~F.~Feng and X.~Q.~Li,
Phys.\ Rev.\ D {\bf 63}, 073006 (2001)
[hep-ph/0012300];
V.~D.~Barger, T.~Han, S.~Hesselbach and D.~Marfatia,
Phys.\ Lett.\ B {\bf 538}, 346 (2002)
[hep-ph/0108261];
S.~K.~Kang and O.~C.~Kong,
hep-ph/0206009;
M.~A.~Diaz, M.~Hirsch, W.~Porod, J.~C.~Romao and J.~W.~F.~Valle,
Phys.\ Rev.\ D {\bf 68}, 013009 (2003)
[hep-ph/0302021].

\bibitem{ghprl}
Y.~Grossman and H.~E.~Haber,
Phys.\ Rev.\ Lett.\  {\bf 78}, 3438 (1997)
[hep-ph/9702421].

\bibitem{HKK}
M.~Hirsch, H.~V.~Klapdor-Kleingrothaus and S.~G.~Kovalenko,
Phys.\ Lett.\ B {\bf 398}, 311 (1997)
[hep-ph/9701253];
M.~Hirsch, H.~V.~Klapdor-Kleingrothaus and S.~G.~Kovalenko,
hep-ph/9701273; 
M.~Hirsch, H.~V.~Klapdor-Kleingrothaus and S.~G.~Kovalenko,
Phys.\ Rev.\ D {\bf 57}, 1947 (1998)
[hep-ph/9707207];
M.~Hirsch, H.~V.~Klapdor-Kleingrothaus, S.~Kolb and S.~G.~Kovalenko,
Phys.\ Rev.\ D {\bf 57}, 2020 (1998).


\bibitem{ghrpv}
Y.~Grossman and H.~E.~Haber,
Phys.\ Rev.\ D {\bf 59}, 093008 (1999)
[hep-ph/9810536].

\bibitem{ghtalk}
Y.~Grossman and H.~E.~Haber,
hep-ph/9906310.



\bibitem{Dav-Los}
S.~Davidson and M.~Losada,
JHEP {\bf 0005}, 021 (2000)
[hep-ph/0005080];
Phys.\ Rev.\ D {\bf 65}, 075025 (2002)
[hep-ph/0010325].


\bibitem{numerical}
A.~Abada, S.~Davidson and M.~Losada,
Phys.\ Rev.\ D {\bf 65}, 075010 (2002)
[hep-ph/0111332];
A.~Abada, G.~Bhattacharyya and M.~Losada,
Phys.\ Rev.\ D {\bf 66}, 071701 (2002)
[hep-ph/0208009].

\bibitem{Davidson:1999mc}
S.~Davidson, M.~Losada and N.~Rius,
Nucl.\ Phys.\ B {\bf 587}, 118 (2000)
[hep-ph/9911317]. 


\bibitem{biloop}
F.~Borzumati and J.~S.~Lee,
Phys.\ Rev.\ D {\bf 66}, 115012 (2002)
[hep-ph/0207184].

\bibitem{Chun}
E.~J.~Chun, D.~W.~Jung and J.~D.~Park,
Phys.\ Lett.\ B {\bf 557}, 233 (2003)
[hep-ph/0211310].

\bibitem{Ibanez:1991pr}
L.~E.~Ibanez and G.~G.~Ross,
Nucl.\ Phys.\ B {\bf 368}, 3 (1992).


\bibitem{sachaellis}
S.~Davidson and J.~R.~Ellis,
Phys.\ Lett.\ B {\bf 390}, 210 (1997)
[hep-ph/9609451];
S.~Davidson and J.~R.~Ellis,
Phys.\ Rev.\ D {\bf 56}, 4182 (1997)
[hep-ph/9702247];
S.~Davidson,
Phys.\ Lett.\ B {\bf 439}, 63 (1998)
[hep-ph/9808425].

\bibitem{Ferr}
J.~Ferrandis,
Phys.\ Rev.\ D {\bf 60}, 095012 (1999)
[hep-ph/9810371].


\bibitem{ghbasis}
Y.~Grossman and H.~E.~Haber,
Phys.\ Rev.\ D {\bf 63}, 075011 (2001)
[hep-ph/0005276].

\bibitem{bgnn}
T.~Banks, Y.~Grossman, E.~Nardi and Y.~Nir,
Phys.\ Rev.\ D {\bf 52}, 5319 (1995)
[hep-ph/9505248].


\bibitem{enrico}
E.~Nardi,
Phys.\ Rev.\ D {\bf 55}, 5772 (1997)
[hep-ph/9610540].

\bibitem{LNS}
M.~Leurer, Y.~Nir and N.~Seiberg,
Nucl.\ Phys.\ B {\bf 398}, 319 (1993)
[hep-ph/9212278];
Y.~Nir and N.~Seiberg,
Phys.\ Lett.\ B {\bf 309}, 337 (1993)
[hep-ph/9304307];
M.~Leurer, Y.~Nir and N.~Seiberg,
Nucl.\ Phys.\ B {\bf 420}, 468 (1994)
[hep-ph/9310320].

\bibitem{FrNi}
C.~D.~Froggatt and H.~B.~Nielsen,
Nucl.\ Phys.\ B {\bf 147}, 277 (1979).

\bibitem{lmmm}
Y.~Grossman and Y.~Nir,
Nucl.\ Phys.\ B {\bf 448}, 30 (1995)
[hep-ph/9502418].

\bibitem{yyy}
Y.~Grossman, Y.~Nir and Y.~Shadmi,
JHEP {\bf 9810}, 007 (1998)
[hep-ph/9808355];
K.~Choi, E.~J.~Chun, K.~Hwang and W.~Y.~Song,
Phys.\ Rev.\ D {\bf 64}, 113013 (2001)
[hep-ph/0107083].

\bibitem{hunter}
J.~F.~Gunion, H.~E.~Haber, G.~L.~Kane and S.~Dawson,
``The Higgs Hunter's Guide'',
[Erratum-ibid.  
hep-ph/9302272].

\bibitem{gh1}
J.~F.~Gunion and H.~E.~Haber,
Nucl.\ Phys.\ B {\bf 272}, 1 (1986)
[Erratum-ibid.\ B {\bf 402}, 567 (1993)].

\bibitem{suspect}
A.~Djouadi, J.~L.~Kneur and G.~Moultaka,
hep-ph/0211331.

\end{thebibliography}
\end{document}